\documentclass[english,aps,amsmath,amssymb,prl,twocolumn]{revtex4}
\usepackage[T1]{fontenc}
\usepackage[utf8]{inputenc}
\usepackage{amsmath}
\usepackage{amssymb}
\usepackage{esint}

\makeatletter
\@ifundefined{textcolor}{}
{%
 \definecolor{BLACK}{gray}{0}
 \definecolor{WHITE}{gray}{1}
 \definecolor{RED}{rgb}{1,0,0}
 \definecolor{GREEN}{rgb}{0,1,0}
 \definecolor{BLUE}{rgb}{0,0,1}
 \definecolor{CYAN}{cmyk}{1,0,0,0}
 \definecolor{MAGENTA}{cmyk}{0,1,0,0}
 \definecolor{YELLOW}{cmyk}{0,0,1,0}
 }

\makeatother

\usepackage{babel}
\begin{document}

\title{Linear response theory for quantum open systems}

\author{J. H. Wei}

\email{wjh@ruc.edu.cn}

\affiliation{Department of Physics, Renmin University of China, Beijing 100872,
P. R. China }

\author{YiJing Yan}

\email{yyan@ust.hk}

\affiliation{Deparment of Chemistry, Hong Kong University of Science and Technology,
Hong Kong, P. R. China}
\begin{abstract}
Basing on the theory of Feynman's influence functional and its hierarchical
equations of motion, we develop a linear response theory for quantum
open systems. Our theory provides an effective way to calculate dynamical
observables of a quantum open system at its steady-state, which can
be applied to various fields of non-equilibrium condensed matter physics.
\end{abstract}
\maketitle
\emph{Introduction}: The linear response theory (LRT) has been widely
used in condensed matter physics since it was derived by Kubo in 1957
\cite{Kub57}. For example, one can calculate the conductivity via
the current-current correlation function directly from the ground
state without applying any realistic bias voltage. However, Kubo's
theory is only valid for an equilibrium closed system (not limited
to quantum one). In recent years, quantum theory for open systems
have attract more and more research interests in molecular electronics,
nanophysics and biophysics, etc. One fundamental issue is whether
we can directly calculate the dynamical observables (e.g. spectra
function) of a static open system (with time-translation-symmetry)
instead of evolving it under some time-dependent field? In this letter,
we address this issue by developing a linear response theory for quantum
open systems (in analogy with Kubo's). Our derivation is based on
the theory of Feynman's influence functional \cite{Fey63} and its
hierarchical equations of motion (HEOM) \cite{Jin08}. 

There exist two correlated questions in developing the LRT for quantum
open systems, 1) how to get the reduced density operator (RDO) of
a static open system; and 2) how to calculate the response to another
probe field? Since question 1 has been well resolved recently via
the HEOM of RDO (and its auxiliary ones) derived by differentiating
Feynman's influence functional \cite{Jin08}, we thus focus on question
2 in the present work.

Before the formal derivation, let us introduce an important character
of influence functional: if a system $S$ simultaneously interacts
with two different environments ($A$ and $B$) and no direct coupling
between them exists at initial conditions, then the total influence
functional $F$ of the system is: $F=F_{A}\cdot F_{B}$ \cite{Fey63}.
It also means the influence phase satisfying: $\Phi(t)=\Phi_{A}(t)+\Phi_{B}(t)$
\cite{Fey63}. When one differentiates the influence functional to
derive the HEOM \cite{Jin08}, those two different phases will linearly
enter into HEOM tie by tie with their simple relationship being maintained
{[}see Eq.\eqref{HEOM_G}{]}. That character provides a practical
solution for question 2 mentioned above.

\ 

\emph{HEOM of propagator}: As the answer of question 1 (and for later
purposes), we first outline the main results of Ref.\onlinecite{Jin08}
in the language of Liouville-space propagator. Let us suppose our
quantum open system composed of the (reduced) system with particle
operator $a_{\mu}^{\sigma}$ ($\sigma=+/-$ corresponding to the creation/annihilation
operator ) and the environment with bath correlation function $C_{\mu\nu}^{\sigma}(t)$
(For fermion bath, $\sigma=+/-$ corresponding to particle transferring
in/out of the system), where $\mu$ ($\nu$) represents the orbital
(site, energy or spin, etc.) index of the system. We then chose $\left\{ \psi\right\} $
as an arbitrary basis set (defining a certain path in path integral
representation) in the system subspace, and $\boldsymbol{\psi}\equiv\left\{ \psi,\psi'\right\} $
for short of the two paths in the theory of influence functional \cite{Fey63}.
Now we can write the HEOM of the influence functional in Liouville
space as {[}$j=\{\mu\sigma\}$, $\bar{j}=\{\mu\bar{\sigma}\}${]}
\footnote{
\[
\mathcal{A}_{j}\equiv a_{\mu}^{\sigma}[\psi(t)]+a_{\mu}^{\sigma}[\psi'(t)].
\]
}%
\footnote{
\[
\tilde{\mathcal{C}}_{j}\equiv\sum_{\nu}C_{\mu\nu}^{\sigma}(t=0)a_{\nu}^{\sigma}[\psi(t)]-\sum_{\nu}C_{\nu\mu}^{\bar{\sigma}}(t=0)a_{\nu}^{\sigma}[\psi'(t)].
\]
}

\begin{equation}
\partial_{t}\mathcal{F}_{\mathbf{j}}^{(n)}=\tilde{\mathcal{F}}_{\mathbf{j}}^{(n)}-i\sum_{k=1}^{n}(-1)^{n-k}\tilde{\mathcal{C}}_{j_{k}}\mathcal{F}_{\mathbf{j}_{k}}^{(n-1)}-i\sum_{j}\mathrm{'}\mathcal{A}_{\bar{j}}\mathcal{F}_{\mathbf{j}j}^{(n+1)},\label{HEOM_F}
\end{equation}
where the sum $\sum'$ runs over all $j\neq j_{k};\; k=1,...,n$.
$\mathcal{F}^{(0)}$ is the superoperator form of the ordinary influence
functional $F$, and $\mathcal{F}^{(n\geqslant1)}$ the auxiliary
ones. The expressions of auxiliary influence functionals appearing
in the rhs of Eq.\eqref{HEOM_F} are %
\footnote{$\mathcal{B}_{j}=-i\sum_{\nu}\int_{t_{0}}^{t}d\tau\, C_{\mu\nu}^{\sigma}(t,\tau)a_{\nu}^{\sigma}[\psi(\tau)]+i\sum_{\nu}\int_{t_{0}}^{t}d\tau\, C_{\mu\nu}^{\bar{\sigma}\,\ast}(t,\tau)a_{\nu}^{\sigma}[\psi'(\tau)].$
$\tilde{\mathcal{B}}_{j}$ is similar to $\mathcal{B}_{j}$ but with
$C_{\mu\nu}^{\sigma}/C_{\mu\nu}^{\bar{\sigma}\,\ast}$ replaced by
$\dot{C}_{\mu\nu}^{\sigma}/\dot{C}_{\mu\nu}^{\bar{\sigma}\,\ast}.$%
}

\begin{subequations}{\label{AIF}}

\begin{eqnarray}
 & \tilde{\mathcal{F}}_{\mathbf{j}}^{(n)}\equiv\bigl(\tilde{\mathcal{B}}_{j_{n}}\mathcal{B}_{j_{n-1}}\cdots\mathcal{B}_{j_{1}}+\cdots+\mathcal{B}_{j_{n}}\cdots\mathcal{B}_{j_{2}}\tilde{\mathcal{B}}_{j_{1}}\bigl)\mathcal{F};\\
 & \mathcal{F}_{\mathbf{j}_{k}}^{(n-1)}\equiv\mathcal{B}_{j_{n}}\cdots\mathcal{B}_{j_{k+1}}\mathcal{B}_{j_{k-1}}\cdots\mathcal{B}_{j_{1}}\mathcal{F};\\
 & \mathcal{F}_{\mathbf{j}j}^{(n+1)}\equiv\mathcal{B}_{j}\mathcal{B}_{j_{n}}\cdots\mathcal{B}_{j_{1}}\mathcal{F}.
\end{eqnarray}
\end{subequations}

The auxiliary RDOs, $\rho_{\mathbf{j}}^{(n)}$, can hen be defined
in terms of the auxiliary influence functionals 
\begin{equation}
\rho_{\mathbf{j}}^{(n)}(t)\equiv\mathcal{U}_{\mathbf{j}}^{(n)}(t,t_{0})\rho(t_{0}),\label{EE_AR}
\end{equation}
where $\mathcal{U}_{\mathbf{j}}^{(n)}(t,t_{0})$ is the time-evolution
superoperator 
\begin{equation}
\mathcal{U}_{\mathbf{j}}^{(n)}(\boldsymbol{\psi},t;\boldsymbol{\psi}_{0},t_{0})\equiv\int_{\boldsymbol{\psi}_{0}[t_{0}]}^{\boldsymbol{\psi}[t]}\mathcal{D}\boldsymbol{\psi}e^{iS[\psi]}\mathcal{F}_{\mathbf{j}}^{(n)}[\boldsymbol{\psi}]e^{-iS[\psi']},\label{EOP}
\end{equation}
with $S[\psi]$ being the classical action functional of the system. 

From Eq.\eqref{HEOM_F}, one can derive the HEOM of RDO as shown in
Ref.\onlinecite{Jin08}. Here, we give the HEOM of the reduced Liouville-space
propagator, defined as $\mathcal{G}_{\mathbf{j}}^{(n)}(t-t_{0})\equiv\mathcal{U}_{\mathbf{j}}^{(n)}(t,t_{0})$
for the time-translation-invariant system 

\begin{subequations}{\label{HEOM_G}}

\begin{eqnarray}
 & \dot{\mathcal{G}}(t)=-[i\mathcal{L}_{s}+i{\cal L}_{sf}(t)]\mathcal{G}(t)-i\sum_{j}\mathcal{A}_{\bar{j}}\mathcal{G}_{j}^{(1)}(t)\text{;}\\
 & \dot{\mathcal{G}}_{\mathbf{j}}^{(n)}(t)=-[i\mathcal{L}_{s}+i{\cal L}_{sf}(t)]\mathcal{G}_{\mathbf{j}}^{(n)}(t)+\tilde{\mathcal{G}}_{\mathbf{j}}^{(n)}(t)-i\sum_{k=1}^{n}\nonumber \\
 & \times(-1)^{n-k}\tilde{\mathcal{C}}_{j_{k}}\mathcal{G}_{\mathbf{j}_{k}}^{(n-1)}(t)-i\sum_{j}\mathrm{'}\mathcal{A}_{\bar{j}}\mathcal{G}_{\mathbf{j}j}^{(n+1)}(t).
\end{eqnarray}
\end{subequations}with the initial condition being $\mathcal{G}_{\mathbf{j}}^{(n)}(t=0)=\delta_{n0}$.
In Eq.\ref{HEOM_G}, $\mathcal{L}_{s}$ is the Liouville operator
of the system, $\mathcal{L}_{s}...\equiv\left[H_{s},\;...\right]$,
while ${\cal L}_{sf}(t)$ is that of arbitrary external time-dependent
field (e.g. the probe field in the LRT). 

\ 

\emph{HEOM space} : We now define the HEOM linear space that can be
seen as an extension of Liouville space. In the former the basic element
is no longer a operators as in the latter, but is $(N+1)$ dimensional
super-vector constituted by the operator and its auxiliary ones. Let
us take the RDO as an example, which is extended as the HEOM-space
RDO, i.e.

\begin{equation}
\boldsymbol{\rho}(t)\equiv\{\rho(t),\:\rho_{\mathbf{j}}^{(n)}(t)[n=1,2...N]\}.\label{HEOMS_R}
\end{equation}
According to Eqs.\eqref{EE_AR} and \eqref{HEOM_G}, the time-evolution
of $\boldsymbol{\rho}(t)$ is determined by the HEOM-space reduced
propogator $\hat{\boldsymbol{\mathcal{G}}}(t,t_{0})$ 

\begin{equation}
\boldsymbol{\rho}(t)=\hat{\boldsymbol{\mathcal{G}}}(t,t_{0})\boldsymbol{\rho}(t_{0}).\label{HEOMS_RT}
\end{equation}
In HEOM space, Eq.\eqref{HEOM_G} can be shorted as

\begin{equation}
\partial\hat{\boldsymbol{\mathcal{G}}}(t,t_{0})/\partial t=-\hat{\boldsymbol{\mathcal{\varLambda}}}(t)\hat{\boldsymbol{\mathcal{G}}}(t,t_{0}),\label{HEOM_PG}
\end{equation}
where $\hat{\boldsymbol{\mathcal{\varLambda}}}(t)$ is the main time-evolution
superoperator acting on super-vectors in HEOM space, which determines
the equation of motion (EOM) of the propagator as well as the density
operator

\begin{equation}
\dot{\boldsymbol{\rho}}(t)=-\hat{\boldsymbol{\mathcal{\varLambda}}}(t)\boldsymbol{\rho}(t).\label{HEOM_RT1}
\end{equation}
Although the concrete form of $\hat{\boldsymbol{\mathcal{\varLambda}}}(t)$
can not be written out separately from Eq.\eqref{HEOM_PG}, one can
easily work out its effect from Eq.\eqref{HEOM_G}

Generalizing above definitions, any operator in Hilbert space now
can be expanded to a super-vector in HEOM space, which obeys similar
EOM as Eq.\eqref{HEOM_RT1},

\begin{equation}
\boldsymbol{A}(t)\equiv\{A,\: A_{\mathbf{j}}^{(n)}(n=1,2...N)\}.\label{HEOM_A}
\end{equation}
The inner product of two vectors $\boldsymbol{A}$ and $\boldsymbol{B}$
in HEOM space is defined as

\begin{equation}
\left\langle \left\langle \boldsymbol{A}|\boldsymbol{B}\right\rangle \right\rangle \equiv\left\langle \left\langle A|B\right\rangle \right\rangle +\sum_{n=1}^{N}\left\langle \left\langle A_{\mathbf{j}}^{(n)}|B_{\mathbf{j}}^{(n)}\right\rangle \right\rangle ,\label{HEOM_IP}
\end{equation}
where $\left\langle \left\langle A|B\right\rangle \right\rangle \equiv\mathrm{Tr}\left[A^{+}B\right]$
being the standard definition of inner product in Liouville space.
Similarly, $\left\langle \left\langle A_{\mathbf{j}}^{(n)}|B_{\mathbf{j}}^{(n)}\right\rangle \right\rangle \equiv\mathrm{Tr}\left\{ \left[A_{\mathbf{j}}^{(n)}\right]^{+}B_{\mathbf{j}}^{(n)}\right\} $.

\ 

\emph{LRT in HEOM space}: We are now on the position to develop the
linear response theory in HEOM space. Suppose a variable $A$ of the
system $S$ is perturbed by a weak probe field $\epsilon_{pr}(t)$,
which linearly couples to the system via its another variable $B$ 

\begin{equation}
H_{pr}(t)=-B\epsilon_{pr}(t).\label{LRT_HPR}
\end{equation}
For simplicity, we suppose both $A$ and $B$ are Hermitian operators
(extending to non-Hermitian cases is straightforward). The change
of the expected value of $A$ caused by the disturbance of $\epsilon_{pr}(t)$
is

\begin{equation}
\delta\bar{A}(t)=\mathrm{Tr}\left[A\delta\rho(t)\right]=\left\langle \left\langle \boldsymbol{A}(0)|\delta\boldsymbol{\rho}(t)\right\rangle \right\rangle ,\label{LRT_DA}
\end{equation}
where $\boldsymbol{A}(0)$ and $\delta\boldsymbol{\rho}(t)$ respectively
denote the expansion of $A$ and $\delta\rho(t)$ in HEOM space, i.e.

\begin{subequations}\label{LRT_ADR}

\begin{eqnarray}
\boldsymbol{A}(0) & = & \{A,\:\boldsymbol{0}\};\\
\delta\boldsymbol{\rho}(t) & \equiv & \{\delta\rho(t),\:\delta\rho_{\mathbf{j}}^{(n)}(n=1,2...N)\}.
\end{eqnarray}

\selectlanguage{english}%
\end{subequations}

We then apply the first order perturbation in the EOM of RDO, 

\begin{equation}
\boldsymbol{\rho}(t)=\boldsymbol{\rho}_{s}(t)+\delta\boldsymbol{\rho}(t)=\hat{\boldsymbol{\mathcal{G}}}(t,\tau)\boldsymbol{\rho}(\tau).\label{LRT_1STR}
\end{equation}
By separating the superoperator $\hat{\boldsymbol{\mathcal{\varLambda}}}(t)$
as $\hat{\boldsymbol{\mathcal{\varLambda}}}(t)=\hat{\boldsymbol{\mathcal{\varLambda}}}_{s}+\hat{\boldsymbol{\mathcal{\varLambda}}}_{pr}(t)$,
and inserting the Dyson's equation in HEOM space%
\footnote{The validity of Dyson's equation in HEOM space is natural. One can
directly verify Eq.\eqref{LRT_DY} by inserting it into Eq.\eqref{HEOM_G}
and letting ${\cal L}_{sf}(t)={\cal L}_{pr}(t)$.%
} into Eq.\ref{LRT_1STR}

\begin{equation}
\hat{\boldsymbol{\mathcal{G}}}(t,\tau)=\hat{\boldsymbol{\mathcal{G}}}_{s}(t-\tau)-\int_{\tau}^{t}d\tau'\hat{\boldsymbol{\mathcal{G}}}(t,\tau')\hat{\boldsymbol{\mathcal{\varLambda}}}_{pr}(\tau')\hat{\boldsymbol{\mathcal{G}}}_{s}(\tau'-\tau),\label{LRT_DY}
\end{equation}
we arrive at

\begin{equation}
\delta\boldsymbol{\rho}(t)=-\int_{0}^{t}d\tau\hat{\boldsymbol{\mathcal{G}}}(t,\tau)\hat{\boldsymbol{\mathcal{\varLambda}}}_{pr}(\tau)\boldsymbol{\rho}_{s}(\tau).\label{LRT_1STDR}
\end{equation}
In above derivation, we have used $\boldsymbol{\rho}_{s}(t)=\hat{\boldsymbol{\mathcal{G}}}_{s}(t-\tau)\boldsymbol{\rho}_{s}(\tau)$
and initial conditions $\delta\boldsymbol{\rho}(\tau=0)=0;\boldsymbol{\quad\rho}(\tau=0)=\boldsymbol{\rho}_{s}(\tau=0)$. 

To proceed, we define the time-independent HEOM-space superoperator
$\hat{\boldsymbol{\mathcal{B}}}$ as

\begin{equation}
\hat{\boldsymbol{\mathcal{B}}}\equiv i\hat{\boldsymbol{\mathcal{\varLambda}}}_{pr}(t)/\epsilon_{pr}(t),\label{LRT_DEFB}
\end{equation}
\foreignlanguage{english}{whose action can be determined from $\hat{\boldsymbol{\mathcal{\varLambda}}}_{pr}(t)\boldsymbol{A}=i[H_{pr}(t),\;\boldsymbol{A}]$
{[}see Eq.\eqref{HEOM_G}{]} as}

\begin{equation}
\hat{\boldsymbol{\mathcal{B}}}\boldsymbol{\rho}=[B,\;\boldsymbol{\rho}].\label{LRT_ACTB}
\end{equation}

\selectlanguage{english}%
Inserting Eq.\eqref{LRT_ACTB} into \eqref{LRT_1STDR}, we finally
get

\begin{equation}
\delta\bar{A}(t)=i\int_{0}^{t}d\tau\left\langle \left\langle \boldsymbol{A}(0)|\hat{\boldsymbol{\mathcal{G}}}(t,\tau)\hat{\boldsymbol{\mathcal{B}}}|\boldsymbol{\rho}(\tau)\right\rangle \right\rangle \epsilon_{pr}(\tau),\label{LRT_A2}
\end{equation}
\foreignlanguage{english}{from which we are ready to define the response
function in HEOM space as}

\begin{equation}
\chi_{AB}(t,\tau)\equiv i\left\langle \left\langle \boldsymbol{A}(0)|\hat{\boldsymbol{\mathcal{G}}}(t,\tau)\hat{\boldsymbol{\mathcal{B}}}|\boldsymbol{\rho}(\tau)\right\rangle \right\rangle .\label{LRT_Resf1}
\end{equation}
\foreignlanguage{english}{Obviously, the definition of Eq. \eqref{LRT_Resf1}
is quite general. For a static quantum open system satisfying time-translation-symmetry,
$\boldsymbol{\rho}(\tau)\rightarrow\boldsymbol{\rho}_{eq}(T)\equiv\{\rho_{eq},\:\rho_{eq}^{(n)}[n=1,2...N]\}$
and $\hat{\boldsymbol{\mathcal{G}}}(t,\tau)\rightarrow\hat{\boldsymbol{\mathcal{G}}}_{s}(t-\tau)$,
which leads to following definition of the response function }

\begin{equation}
\chi_{AB}(t)\equiv i\left\langle \left\langle \boldsymbol{A}(0)|\hat{\boldsymbol{\mathcal{G}}}_{s}(t)\hat{\boldsymbol{\mathcal{B}}}|\boldsymbol{\rho}_{eq}(T)\right\rangle \right\rangle .\label{LRT_resf2}
\end{equation}

\selectlanguage{english}%
The correlation function in HEOM space can be similarly defined as
\footnote{In our theory, the fluctuation-dissipation theorem is naturally established.
See Eq.\eqref{SF_FDT}.%
}

\begin{equation}
\widetilde{C}_{AB}(t)\equiv\left\langle \left\langle \boldsymbol{A}(0)|\hat{\boldsymbol{\mathcal{G}}}_{s}(t)\boldsymbol{B}|\boldsymbol{\rho}_{eq}(T)\right\rangle \right\rangle ,\label{LRT_corf1}
\end{equation}
\foreignlanguage{english}{where }$\boldsymbol{B}\boldsymbol{\rho}_{eq}(T)=\{B\rho_{eq},\: B\rho_{eq}^{(n)}(n=1,2...N)\}$.

\selectlanguage{english}%
For practical calculation, we can rewrite Eq.\eqref{LRT_resf2} as

\begin{equation}
\chi_{AB}(t)=\left\langle \left\langle \boldsymbol{A}(0)|\boldsymbol{\sigma}(t)\right\rangle \right\rangle =\mathrm{Tr}\left[A^{+}\sigma(t)\right],\label{LRT_resf3}
\end{equation}
\foreignlanguage{english}{where}

\begin{subequations}\label{LRT_sig}

\begin{eqnarray}
\boldsymbol{\sigma}(t) & = & \hat{\boldsymbol{\mathcal{G}}}_{s}(t)\boldsymbol{\sigma}(0);\\
\boldsymbol{\sigma}(0) & = & i\hat{\boldsymbol{\mathcal{B}}}\boldsymbol{\rho}_{eq}(T)\nonumber \\
 & = & \{i[B,\:\rho_{eq}],\: i[B,\:\rho_{eq}^{(n)}](n=1,2...N)\}.
\end{eqnarray}

\end{subequations}

\selectlanguage{english}%
Similarly, if we set $\boldsymbol{\sigma}(0)=\boldsymbol{B}\boldsymbol{\rho}_{eq}(T)$,
Eq.\eqref{LRT_corf1} can be rewritten as an easier handling form

\begin{equation}
\widetilde{C}_{AB}(t)=\left\langle \left\langle \boldsymbol{A}(0)|\boldsymbol{\sigma}(t)\right\rangle \right\rangle =\mathrm{Tr}\left[A^{+}\sigma(t)\right].\label{LRT_corf2}
\end{equation}

\ 

\selectlanguage{english}%
\emph{Spectra function in HEOM space}: In principle, one can calculate
the dynamical correlation of any two system-operators ($A$ and $B$)
via the HEOM-space LRT. One typical example is the spectra function
that plays important role in many body physics. In what follows, we
will demonstrate how to obtain the (reduced) spectra function of a
quantum open system. The spectra function $J_{AB}(\omega)$ directly
relates to the imaginary part of the retarded single-particle Green's
function $G_{AB}^{r}(t)$ in the form of

\begin{subequations}\label{SF_defGJ}

\begin{eqnarray}
G_{AB}^{r}(t) & \equiv & -i\theta(t)\left\langle \left\{ A(t),\: B\right\} \right\rangle ;\\
J_{AB}(\omega) & \equiv & -\frac{1}{\pi}\mathbf{\mathrm{Im}}\left[G_{AB}^{r}(\omega)\right].
\end{eqnarray}
\end{subequations}Please be noted that the Green's function in our
theory is defined for two arbitrary operators, which can reduce to
the one in textbooks by setting $A=a$ and $B=a^{+}$.

For fermion, $G_{AB}^{r}(t)$ can not be obtained directly from $\chi_{AB}(t)$
due to the anti-communication relation in the former but communication
one in the latter. Fortunately, we can get $G_{AB}^{r}(t)$ from correlation
function $\widetilde{C}_{AB}(t)$, i.e.

\begin{eqnarray}
G_{AB}^{r}(t) & = & -i\theta(t)\left\langle \left\{ A(t),\: B\right\} \right\rangle \nonumber \\
 & = & -i\theta(t)\left[\widetilde{C}_{AB}(t)+\widetilde{C}_{BA}(-t)\right].\label{SF_GR1}
\end{eqnarray}
To proceed, we introduce the general spectra function $C_{AB}(\omega)\equiv\frac{1}{2}\int_{-\infty}^{+\infty}dt\, e^{i\omega t}\widetilde{C}_{AB}(t)$,
which satisfies the detailed-balance-relation $C_{BA}(-\omega)=e^{-\beta\omega}C_{AB}(\omega)$.
After some algebra, we obtain

\begin{equation}
J_{AB}(\omega)=\frac{1}{\pi}\left(1+e^{-\beta\omega}\right)C_{AB}(\omega),\label{SF_FDT}
\end{equation}
which is obviously the fluctuation-dissipation theorem in HEOM space.

Since in Eq.\eqref{LRT_corf1} only $\hat{\boldsymbol{\mathcal{G}}}_{s}(t)$
is the function of time, we have

\begin{eqnarray}
C_{AB}(\omega) & = & \left\langle \left\langle \boldsymbol{A}(0)|\hat{\boldsymbol{\mathcal{G}}}_{s}(\omega)\boldsymbol{B}|\boldsymbol{\rho}_{eq}(T)\right\rangle \right\rangle ,\label{SF_CW}
\end{eqnarray}
where $\hat{\boldsymbol{\mathcal{G}}}_{s}(\omega)$ is the Fourier
transform of Eq.\eqref{HEOM_G}. Thus $J_{AB}(\omega)$ can be calculated
from 

\begin{equation}
J_{AB}(\omega)=\frac{1}{2\pi}\left(1+e^{-\beta\omega}\right)\left\langle \left\langle \boldsymbol{A}(0)|\hat{\boldsymbol{\mathcal{G}}}_{s}(\omega)\boldsymbol{B}|\boldsymbol{\rho}_{eq}(T)\right\rangle \right\rangle .\label{SF_JW1}
\end{equation}
More concretely, if choosing $\boldsymbol{\sigma}(0)=\boldsymbol{B}\boldsymbol{\rho}_{eq}(T)=\{B\rho_{eq},\: B\rho_{eq}^{(n)}(n=1,2...N)\}$,
then we have $\boldsymbol{\sigma}(\text{\ensuremath{\omega}})=\hat{\boldsymbol{\mathcal{G}}}_{s}(\omega)\boldsymbol{\sigma}(0)$
and 

\begin{eqnarray}
J_{AB}(\omega) & = & \frac{1}{2\pi}\left(1+e^{-\beta\omega}\right)\left\langle \left\langle \boldsymbol{A}(0)|\boldsymbol{\sigma}(\omega)\right\rangle \right\rangle \nonumber \\
 & = & \frac{1}{2\pi}\left(1+e^{-\beta\omega}\right)\mathrm{Tr}\left[A^{+}\sigma(\omega)\right].\label{SF_JW2}
\end{eqnarray}
Eq.\eqref{SF_JW2} is the main formula to calculate HEOM-space spectra
function in our theory. Before that, one must solve equation $\hat{\boldsymbol{\mathcal{\varLambda}}_{s}}\boldsymbol{\rho}_{eq}(T)=0$
to obtain $\boldsymbol{\rho}_{eq}(T)$.

\ 

\emph{Summary}: In summary, we have developed a linear response theory
for quantum open systems on the basis of the theory of Feynman's influence
functional and its hierarchical equations of motion. From our theory,
one can directly calculate the dynamical observables (e.g. spectra
function) of a static open system instead of evolving it under time-dependent
field. It can be applied to non-equilibrium many-body physics, nanophysics,
etc.

\end{document}